# Trapped particle bounds on stimulated scatter in the large k$\lambda_D$ regime

Harvey A. Rose[*]

Los Alamos National Laboratory

Los Alamos, NM 87544

In the strongly damped regime, the convective gain rate for stimulated scatter, $\kappa$, is customarily maximized by requiring that, taken together, the laser light wave and the daughter light and plasma waves, satisfy wavevector and frequency matching, and then $1/\kappa \sim \gamma$, the plasma wave damping rate. If the bounce frequency in the daughter plasma wave is large compared to the trapped particle loss rate, it would seem, based on naïve extrapolation of the work by Zakharov and Karpman (V. E. Zakharov and V. I. Karpman, JETP **16**, 351 (1963)) on decaying, one dimensional (1D) Langmuir waves, that $\kappa$ may be increased indefinitely by increasing the electrostatic wave amplitude, $\phi$, since they calculate that $\gamma$ varies as $\phi^{-3/2}$. However, for a driven plasma wave in a laser speckle—as is appropriate to stimulated Raman scatter in an optically smoothed laser beam in 3D—it has been shown (H. A. Rose and D. A. Russell, Phys. Plasma, **8**, 4784 (2001)) that $\gamma$ varies more slowly, $\propto \phi^{-1/2}$, and asymptotes to a finite value for large $\phi$, when the loss of trapped electrons due to convection out the speckle sides dominates that due to collisions. This behavior, combined with the loss of resonance for $\phi$ too large, leads to a maximum value for $\kappa$ as a function of scattered light frequency and $\phi$, for given laser and plasma parameters. Bounds for stimulated Brillouin scatter gain rate are also obtained. The standard mode-coupling model (MCM) of these scattering processes, when modified to include the trapped particle nonlinear frequency shift, always allows for a propagating plasma wave, and therefore may be qualitatively in error in regimes where the daughter plasma wave loses resonance. A mean field approximation model is proposed which is consistent with the bound on $\kappa$ and agrees with the MCM in the resonant regime, but differs in the non-resonant regime by respecting this fundamental difference in the plasma mode structure. If a plasma, as it evolves, crosses the resonant/non-resonant regime boundary, a model that is cognizant of both regimes is required to avoid a qualitative overestimate of the scatter.

---

[*] har@lanl.gov

## I. INTRODUCTION

Stimulated Raman and Brillouin scatter have long been considered as deleterious to the achievement of inertial confinement fusion[1]. *A priori* attempts at their control have either relied on gain saturation by plasma gradients[2], or given the large plasma scale lengths anticipated for hohlraum targets at the National Ignition Facility[1] (NIF), by large values of Landau damping to keep the gain below the level at which thermal fluctuations may be amplified to finite levels. However, there is a concern that the effective damping may be lower than classical (*i.e.*, Landau damping), either as a result of ohmically modified distribution functions[3], or trapping effects, as evidenced by recent observations[4] of backscatter stimulated Raman scatter (BSRS), in conjunction with stimulated electron acoustic scatter (SEAS). This data strongly suggests that SEAS is beyond the ken of a model that ignores electron-trapping effects on the electron acoustic mode. Since the Bohm-Gross (Langmuir wave) branch and electron acoustic branch of BGK modes merge at short wavelengths ($k\lambda_D \approx 0.53$, where $k$ is the wavenumber of the electron plasma wave and $\lambda_D$ the electron Debye length$=1/k_D$), and since the anticipated NIF parameter regime may straddle this wavenumber regime, both these branches and their coupling need further study.

For the NIF, with a low Z (ionic charge) hohlraum plasma fill— a Helium-Hydrogen mixture—and, at the peak of the laser pulse, an estimated electron temperature, $T_e$, of 5 keV, ohmic effects on the electron distribution function appear to be a correction, and will be ignored here. The primary purpose of this paper is to present estimates as to how large the BSRS and backward stimulated Brillouin scatter (BSBS) spatial gain rate coefficients can get due to trapping effects. The physical arguments behind these estimates indicate that the standard mode-coupling model of these scattering processes, when modified to include the trapped particle nonlinear frequency shift, may be qualitatively in error when the daughter plasma wave loses resonance. An alternative model, which agrees with the former when there is a resonance, is introduced whose validity does not require a resonant plasma wave.

## II. REVIEW of PLASMA WAVE RESPONSE IN THE STRONGLY TRAPPED REGIME

It is initially assumed that the plasma wave response is strictly local: given an external potential, $\mathrm{Re}\,\phi_0 \exp[i(kx - \omega t)]$, with the envelope function $\phi_0$ possibly having slow space and time variations, then the total potential, $\Phi = \mathrm{Re}\,\phi \exp[i(kx - \omega t)]$, is given by $\phi = \phi_0/\varepsilon$, where $\varepsilon$ is the nonlinear dielectric function whose dependence on $k$, $\omega$, and $\phi$ is described below. When the internal component of $\phi$, $\phi_{\mathrm{int}}$, $\phi = \phi_0 + \phi_{\mathrm{int}}$, which is related to plasma charge density fluctuations via Poisson's equation, is determined by electrons, as in the case of SRS and SEAS, then

$$\varepsilon = 1 - \Xi/(k\lambda_D)^2. \tag{1}$$

The nonlinear susceptibility, $\Xi$, is defined as the normalized ratio of the electron density fluctuation to the total potential,

$$\Xi = \frac{\delta n_e/n_e}{e\phi/T_e}. \tag{2}$$

In this expression, the sign of "$e$" and the fact that $\phi$ is complex valued, count. $\delta n_e$ is the complex valued envelope of the density fluctuation at $k$, and $n_e$ is the background density. It differs, in the linear regime, from the usually defined susceptibility by a sign and by removing the explicit factor of $(k\lambda_D)^{-2}$, which now appears in (1).

In the linear regime, $\phi \to 0$, and for a background Maxwellian distribution function, $f_0$, it follows that

$$\Xi \to \Xi_0(\zeta) = Z'(\zeta/\sqrt{2})/2. \tag{3}$$

$\zeta = v/v_e$, and $Z$ is the plasma dispersion function[5]. v is the wave's phase speed, $v=\omega/k$, and $v_e$ is the electron thermal speed. In the nonlinear regime, a relaxation term is added to the Vlasov equation, so that in the wave frame

$$\left(\frac{\partial}{\partial t} + u\frac{\partial}{\partial x} - \frac{e}{m_e}\frac{\partial \Phi}{\partial x}\frac{\partial}{\partial u}\right) f(x,u,t) = -\nu[f(x,u,t) - f_0(u+v)], \tag{4}$$

and now $\Phi$ has no explicit time dependence. $m_e$ is the electron mass. $\nu$ is interpreted as the escape rate of trapped electrons in the omitted transverse dimensions and is estimated by the rate at which a thermal electron traverses a laser speckle (intensity "hot spot" or just "hot spot") width, $F\lambda_0$, namely $v_e/F\lambda_0$, with $F$ the optic $f/\#$ and $\lambda_0$ the laser wavelength. For this model, $\Xi_0$ has a modified argument, $\zeta = v/v_e + i\mu$, $\mu = \nu/kv_e$.

### A. Perturbative evaluation of the real part of $\Xi$

$\Xi$ is evaluated assuming a steady response to time independent $\phi_0$. It depends on three dimensionless arguments, which may be taken as[6] $(v/v_e, |e\phi|/T_e, \mu)$. It has been shown[7], in the strongly trapped regime, $\nu/\omega_b \ll 1$, with $\omega_b$ the bounce frequency ($\omega_b/\omega_p = k\lambda_D\sqrt{e\phi/T_e}$, $\omega_p$ is the electron plasma frequency), that to lowest order in $\phi$ and $\mu$,

$$\mathrm{Re}(\Xi) \equiv \mathrm{Re}(\Xi_0) + \delta\Xi = \mathrm{Re}(\Xi_0) - 1.76 f_0''(v/v_e)\sqrt{e\phi/T_e} + O(\mu^2). \tag{5}$$

The normalization of $f_0$ is now chosen such that $\int f_0(u)du = \int u^2 f_0(u)du = 1$, and it is defined in its rest frame, $\int u f_0 du = 0$, so that, e.g., for the Maxwellian case $f_0(x) = (1/\sqrt{2\pi})\exp(-0.5x^2)$. There is no explicit wavenumber limitation of this approximation's validity, nor of the implied plasma wave frequency shift[8] determined by comparing resonances, i.e., solutions of Re($\varepsilon$)=0, for $\phi$=0 and finite $\phi$. However there are various amplitude constraints. One such constraint[9] is based on ignoring the anharmonic corrections to the self-consistent potential, whose accuracy requires that $|f_0''(v/v_e)|\sqrt{e\phi/T_e} \ll 20(k\lambda_D)^2$. On the nonlinear extension of the Bohm-Gross branch ($v/v_e \approx 1/k\lambda_D$), this is not very demanding. Since (5) is a perturbative result, it must break down for large enough $\phi$, although good accuracy has been attained for $e\phi/T_e$ as large as 1/2, for various special cases considered in [7].

With (5) in hand, one can address the question of nonlinear loss or resonance (LOR). As is plain from figures 1 and 2 in reference [7], as $\phi$ increases, the maximum value of $k\lambda_D$ for which a resonance is possible decreases. Its squared value is the maximum of Re($\Xi$) over v. For $\phi$=0 and Maxwellian $f_0$, it is known[10] to occur at $k\lambda_D \approx 0.53$.

However, $v/v_e$ cannot be too small, or else (4) loses its physical validity, as explained below. This issue does not arise for electron dynamics alone since the use of this expression for $\Xi$, equation (5), is usually in the context of "near BGK modes", traveling wave solutions of (4) with $\phi \gg \phi_0$, for which $v/v_e$ is order unity or greater. If dynamic ions are included, so that ion acoustic BGK modes are possible, then $v/v_e \ll 1$ for these modes, and the electron contribution to $\Xi$ must be reconsidered.

Equation (4) is a one-dimensional surrogate for a full three-dimensional (3D) model. In 3D, convection out the sides of a speckle in which $\Phi$ is localized, is explicitly included. Instead of a relaxation term on the right hand side (rhs) of (4), the corresponding 3D model has a "0". If v vanishes, and at infinity, $f$ approaches $f_0$, a Maxwellian, then on general principles, the equilibrium solution is given by $-\ln f \sim \frac{1}{2} m u^2 + \Phi(\mathbf{x})$, and $\Xi$ is given by $\left[\exp(-e\Phi/T_e) - 1\right] / e\Phi/T_e = -1 + O(\Phi)$. Gone is the singular dependence on $\phi$ as $\phi \to 0$, the $\sqrt{\phi}$ term in (5), which obtains for finite v. This dependence reflects the mismatch—whose resolution occurs across the separatrix between trapped and passing particle orbits—between $f_0$, which in the frame of the wave depends on the kinetic energy based on the relative velocity, $\sim (u+v)^2$, while $f$, which is close to a BGK mode if $v/\omega_b \ll 1$, is based on the total energy, whose kinetic component $\sim u^2$.

For small $v/v_e$, it is well known that the linear electron response has an imaginary part, resulting in a small contribution to the damping of ion acoustic waves, which is ignored.

**B. Perturbative evaluation of the imaginary part of $\Xi$**

Its perturbative evaluation, Im($\Xi_{per}$), is given by[7]

$$\text{Im}(\Xi_{per}) = 6.2 f_0'(v/v_e) \frac{\mu}{\sqrt{e\phi/T_e}} + \mu\Delta(v/v_e), \tag{6}$$

in the strongly trapped regime. $\Delta$ is given below in (8). Unlike the real part, the imaginary part of $\Xi$, when evaluated in the strongly trapped regime, does not go over to its linear value for small $\phi$, which for small $\mu$ is given by

$$\text{Im}(\Xi_0) = -2(k\lambda_D)^2 v_{\text{Landau}}/\omega_p = \pi f_0'(v/v_e). \tag{7}$$

$v_{\text{Landau}}$ is the classic Landau damping rate. As a result, $\Xi$ may not be generally expressed as $\Xi_0$+perturbation, instead, equations (5) and (6) may be combined to express the total susceptibility as, $\Xi \approx \text{Re}(\Xi_0) + \delta\Xi + i\,\text{Im}(\Xi_{per})$, which is only valid for $v/\omega_b \ll 1$. An extrapolation of this result to the weakly trapped regime is presented in the next section.

There are two apparent differences between (6) and the result of Zakharov and Karpman (ZK)[11]. First, from dimensional analysis, if the relaxation of the distribution function is diffusive, $D\partial^2 f/\partial v^2$, as considered in ZK, then the natural dimensionless combination for residual damping[12] varies as $D/\phi^{3/2}$, instead of $v/\omega_b \propto v/\sqrt{\phi}$ as in (6). Secondly, and qualitatively more significant, there is the second term on the rhs of (6). In reference [7] it is obtained not as the next term in an expansion in powers of $\phi$ but rather via a mean field approximation with

$$\Delta(x) = 3\int_{u>0} \left[ f_0(x+u) - f_0(x-u) - 2u f_0'(x) \right] \frac{du}{u^3}. \tag{8}$$

Since the magnitude of $\mu\Delta$ is a lower bound[13] for the damping in the strongly trapped regime, it may not, in fact, be a reduction compared to the linear damping given by (7). Unless $\mu \ll \text{Im}(\Xi_0)/\Delta$, there cannot be a large decrease in damping due to trapping.

The graph of Im($\Xi_0$)/$\Delta$, is shown in figure 1 for Maxwellian $f_0$. In this graph, $k\lambda_D$ is related to v/v$_e$ through the linear dispersion relation, $(k\lambda_D)^2 = \text{Re}[\Xi_0(v/v_e)]$, evaluated on the Bohm-Gross branch, *e.g.*, v/v$_e$={2.5, 3.0, 3.5, 4.0} corresponds to $k\lambda_D$={0.51, 0.42, 0.35, 0.29}.

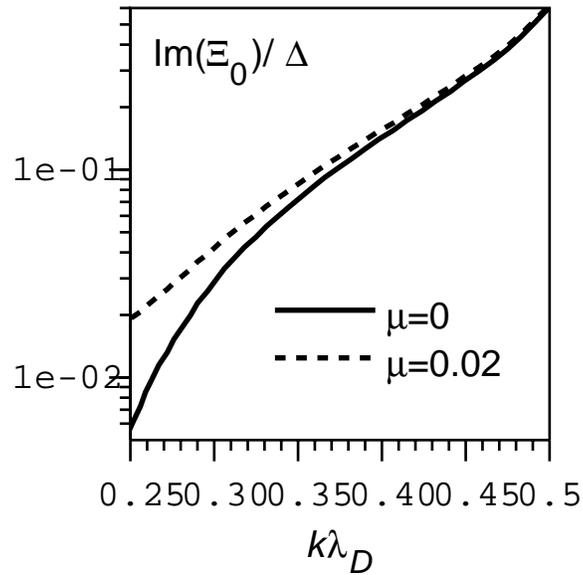

FIG. 1. The normalized escape rate of trapped electrons, $\mu$=$\nu$/$k$v$_e$, must be well below the ordinate of this graph for trapping to effect a significant reduction in damping.

$\nu$ =v$_e$/$F\lambda_0$ implies that $\mu = 1/(F k\lambda_0)$, and for backscatter at low density, $k \approx 2k_0$, one has the estimate $\mu \approx 1/4\pi F$, so that for modest values of $F$, $\mu$ is order 0.01, and $k\lambda_D$ must be greater than about 0.3 for trapping to lead to a large damping reduction. Since the dependence of $\Xi$ on $k$ is through its dependence on $\mu$, $\Xi$ is but weakly dependent on $k\lambda_D$, and therefore $\varepsilon$ depends on $k\lambda_D$ primarily through the explicit factor in (1), which validates the utility of this representation.

## C. Evaluation of the imaginary part of Ξ in transition

Since (6) is singular as $\phi \to 0$, and since for $k\lambda_D \geq 0.53$ it will be shown that the maximum gain rate for BSRS is attained for $\phi \approx 0$, a smooth transition must be imposed on Im(Ξ) as $\phi$ varies between the strongly and weakly trapped regimes. The quasi-linear (QL) approximation has been found[7] to provide accurate values for Ξ, and is simpler to use than the numerically exact methods. The defining equations for QL theory, as applied to the evaluation of Ξ for a coherent potential, are reproduced here from [7] for convenience:

$$f_{QL}(x) - f_0\left(x + \frac{v}{v_e}\right) = \frac{1}{2}(e\phi/T_e)^2 \frac{\partial}{\partial x}\left[\frac{1}{|x - i\mu|^2}\frac{\partial f_{QL}}{\partial x}\right] \tag{9}$$

$$\Xi_{QL} = \int \frac{f_{QL}(x)}{(x - i\mu)^2} dx. \tag{10}$$

In figure 2, various approximations to Im(Ξ) are shown for the case[14] $\mu=0.01$, $v/v_e=3.0$. The curve labeled "GM2", is the negative of a quadratic geometric mean between Im($\Xi_{per}$) and Im($\Xi_0$), defined by

$$\text{GM2}(a,b) = ab/\sqrt{a^2 + b^2}. \tag{11}$$

The classic geometric mean, $ab/(a+b)$, gives a qualitatively inferior fit in the transition region. The smallest value of the ordinate shown, −0.043, is Im($\Xi_0$). It has been found by detailed numerical comparisons that the GM2 fit is quantitatively accurate, over the range of $e\phi/T_e$ which spans the weakly to strongly trapped transition regime, for $0.35 < k\lambda_D < 0.51$, and $\mu=0.01$ and $0.02$. This statement is not meant to imply inaccuracy for other values of these parameters. In summary, the expression used for $\Xi$, which incorporates perturbative evaluations and the GM2 fit, and has numerically proven to be valid in both the weakly and strongly trapped regime, is

$$\Xi(v/v_e, e\phi/T_e, \mu) = \text{Re}(\Xi_0) + \delta\Xi - i\text{GM2}\left[\text{Im}(\Xi_{per}), \text{Im}(\Xi_0)\right]. \tag{12}$$

$\Xi_0$ has argument $v/v_e + i\mu$, $\delta\Xi$ is given by (5) and $\text{Im}(\Xi_{per})$ by (6).

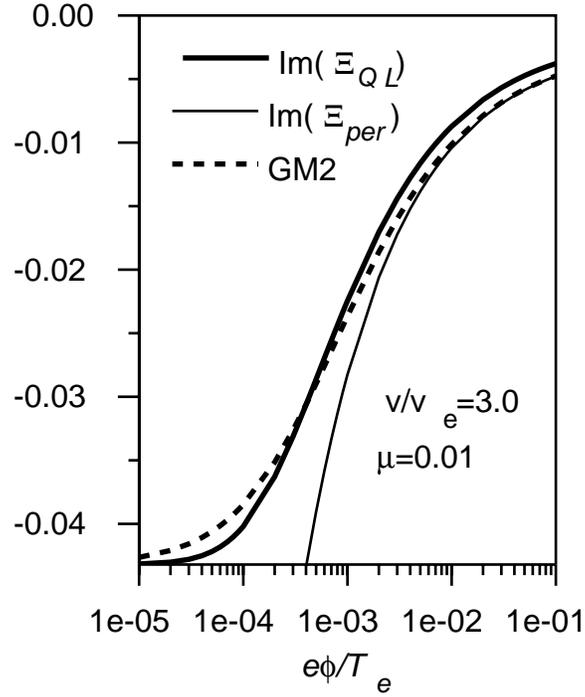

Fig. 2. Various approximations to the imaginary part of the nonlinear susceptibility, $\Xi$, for $v/v_e=3.0$ and $\mu=0.01$.

# III. GAIN RATE UPPER BOUND AND RELATION TO THE MODE COUPLING and MEAN FIELD MODELS

As suggested by Cohen and Kauffman[15], given $\varepsilon(\phi)$, the standard three-wave model for stimulated scatter may be converted into a nonlinear model. Let the electric field of the laser light be represented by $E_0 = \mathrm{Re}\, E_0 \exp[i(k_0 z - \omega_0 t)]$, and for BSRS, the scattered light by $E_{srs} = \mathrm{Re}\, E_{srs} \exp[i(k_{srs} z - \omega_{srs} t)]$. $k_0$ and $\omega_0$ are taken positive so that the laser light propagates in the positive "z" direction, while the scattered light is assumed to propagate in the opposite sense. The ponderomotive potential has a high frequency part, $\Phi_0$, which is a source of Langmuir waves, $\Phi_0 = \mathrm{Re}\, \phi_0 \exp[i(kz - \omega t)]$, with

$$\phi_0 = -\frac{i\mathrm{v}_{osc}}{2\omega_{srs}} E_{srs}^*, \tag{13}$$

and $k + k_{srs} = k_0$, $\omega + \omega_{srs} = \omega_0$. $\mathrm{v}_{osc}$ is the electron oscillating electron velocity, $\mathrm{v}_{osc} = ieE_0/m_e\omega_0$. As in [15] (though spatial transport was not allowed), slow variation of $\phi_0$ allows the Langmuir wave response to be approximately given by

$$\left[\frac{\partial}{\partial t} + \mathrm{v}_L \frac{\partial}{\partial z} - \frac{i\varepsilon}{\partial \varepsilon/\partial \omega}\right]\phi = -\frac{1}{\partial \varepsilon/\partial \omega} \frac{\mathrm{v}_{osc}}{2\omega_{srs}} E_{srs}^* = -\frac{i}{\partial \varepsilon/\partial \omega} \phi_0, \tag{14}$$

with $\mathrm{v}_L$ the Langmuir wave group velocity, $\mathrm{v}_L = -(\partial \varepsilon/\partial k)/(\partial \varepsilon/\partial \omega)$, and $\partial \varepsilon/\partial \omega \approx 2/\omega_p$ for large phase velocities. Limitations of this model associated with LOR are discussed later. Since $\varepsilon$ is given by (1) and (12), omitted from (14) are variations of $\varepsilon$ due to changes in the background density and temperature, which would otherwise detune the instability. Therefore estimates of gain thereby obtained are an upper bound. Dispersive effects are ignored. The model is completed with the standard[16]

$$\left(\frac{\partial}{\partial t} + \mathrm{v}_{srs} \frac{\partial}{\partial z}\right) E_{srs} = -e\pi \mathrm{v}_{osc} \delta n_e^* = -\frac{1}{4} k^2 \mathrm{v}_{osc} \phi_{int}^* = -\frac{1}{4} k^2 \mathrm{v}_{osc} (\phi - \phi_0)^*, \tag{15}$$

which together with (13) implies

$$\left(\frac{\partial}{\partial t} + \mathrm{v}_{srs} \frac{\partial}{\partial z} - i\frac{k^2}{8\omega_{srs}} |\mathrm{v}_{osc}|^2\right) E_{srs} = -\frac{1}{4} k^2 \mathrm{v}_{osc} \phi^*, \tag{16}$$

and $v_{srs} \approx -c$, the speed of light, at low densities. Except for a change in notation, "*srs*"→"*sbs*", (13) and the first equality of (15) are also valid for BSBS. Diffraction, refraction and light wave collisional damping have been ignored in the interest of obtaining a gain upper bound. Equations (14) and (16) (and the corresponding equation for $E_0$) constitute the mode coupling model[17] (MCM).

In (14), the wavenumber and frequency arguments of $\varepsilon$, $k$ and $\omega$, are fixed but not independent. Equation (15) follows from the full wave equation only if $k_{srs}$ and $\omega_{srs}$ are related through the light wave dispersion relation which, at low density, $\omega_0 \approx k_0 c$, $\omega_{srs} \approx |k_{srs}|c$, implies that[18]

$$\omega \approx c(2k_0 - k). \tag{17}$$

Since stimulated scatter tends to select the most responsive modes, the canonical choice of $k$ is such that the Langmuir wave (or electron acoustic wave) is resonant[19], Re($\varepsilon$)=0 for $\phi$=0. More generally, one might choose $k$ which maximizes the linear convective gain rate, as given below in (19).

The gradient linearization, equation (14), is, at each spatial location, about the local value of $|\phi|$. In general, the evaluation of $\partial\varepsilon/\partial k$ and $\partial\varepsilon/\partial\omega$ yields a complex valued $v_L$, but at large phase velocities the real part is dominant and the complex part is ignored[20]. This breaks down near LOR where the electron acoustic and Langmuir wave branches merge and $\text{Re}(\partial\varepsilon/\partial\omega) \to 0$. This regime is otherwise problematic in the framework of the MCM because one pair (plasma wave and scattered light wave) is required for each branch. Since their properties approach each other near LOR, one expects their amplitudes to be comparable so that neither may be ignored. Unless their bounce frequencies are small compared to their frequency separation, their effect on $\varepsilon$ will not be a mere superposition. The dependence of $\varepsilon$ on their amplitudes and phases is as yet unknown in this regime.

**A. Gain rate upper bound**

In the strong damping regime, one ignores spatial transport of the plasma wave, but retains the time derivative so that gain as a function of frequency may be obtained. It is now assumed that if all the gain were lumped into a single, optimally chosen frequency, this would give an upper bound, so that for this purpose, (14) is replaced by

$$\varepsilon \phi = \phi_0, \tag{18}$$

and correspondingly the time derivative in (16) is dropped. Equations (16) and (18) then imply, at low density, that the normalized BSRS *amplitude* convective gain rate, $\kappa^{srs}_{norm}$, is

$$\kappa^{srs}_{norm} \equiv \frac{\kappa^{srs}}{k_0}\left(\frac{v_e}{v_{osc}}\right)^2 \approx \frac{1}{8}\frac{n_e}{n_c}\left(\frac{k}{k_D}\right)^2 \frac{|\text{Im}\varepsilon|}{|\varepsilon|^2}. \tag{19}$$

Except for the dependence of $\varepsilon$ on $\phi$, it may be seen that this is the standard expression for $\kappa^{srs}$, the basic BSRS gain rate at low density, once it is recalled that for $\phi = 0$, and $k$ chosen to be at resonance, $|\varepsilon|^2/|\text{Im}\varepsilon| \to |\text{Im}\varepsilon| = 2\nu_{Landau}/\omega_p$. The object now is, for given plasma and laser parameters, to maximize $\kappa^{srs}$ over $k$ and $\phi$. Also, $k_0/k$ (required to obtain $v = \omega/k$) can be re-expressed in terms of $k/k_D$ and $k_0/k_D \approx (v_e/c)\sqrt{n_c/n_e}$, so that $\kappa^{srs}/k_0$ depends on the laser and plasma parameters, $n_e/n_c$, $v_{osc}/v_e$, $T_e$, besides $k/k_D$ and $e\phi/T_e$. It also depends on $\nu/\omega_p$.

Since the dependence of $\kappa^{srs}$ on laser intensity, $I \sim (v_{osc}/v_e)^2$, is simply multiplicative, one must take care, once its maximum value is determined, that the intensity is not so large as to violate the strong damping approximation: it is required that[21] $v_L \kappa^{srs} \ll \omega_p |\text{Im}\varepsilon|$. If this inequality is violated, the error in the use of the strong damping approximation may be large: a generalized absolute (self-sustaining) instability might be induced by trapping effects, even if the intensity is well below the linear absolute threshold.

**B. Comparison with time dependent solutions**

Numerical solutions of the MCM are presented to show the utility and relevance of the upper bound theory. $\phi$ ($E_{srs}$) is taken to vanish at the left (right) side of the simulation region since $v_L>0$ ($v_{srs}<0$). $v_{osc}$ is assumed constant, as pump depletion should not lead to an increase of reflectivity. A thermal source is added to the rhs of (14), which is a delta correlated complex random field whose amplitude is chosen to yield the appropriate level of scatter for small laser intensity. For the illustrative choice of plasma parameters, $n_e/n_c = 0.05$, $1/2\mu$m light, and $T_e=1$keV, this noise results in a reflectivity which increases linearly with distance at the rate $\approx 5\text{E}-12/\mu$m. $\nu/\omega_p=0.005$, a typical value, and the dependence of the maximum gain rate on this parameter is briefly studied later. Various cases are now considered with $n_e/n_c$ and laser intensity as parameters.

For $n_e/n_c=0.05$, the calculated maximum value of $\kappa_{norm}^{srs}$, $\left(\kappa_{norm}^{srs}\right)_{max}=0.040$ at $k/k_D=0.332$ with $e\phi/T_e=0.24$. This is at the top of a broad maximum with regard to variations in $\phi$, as shown in figure 3 (solid curve). The $n_e/n_c=0.03$ case is also shown. Each of these curves is generated by maximizing $\kappa_{norm}^{srs}$ over $k\lambda_D$ for given $\phi$.

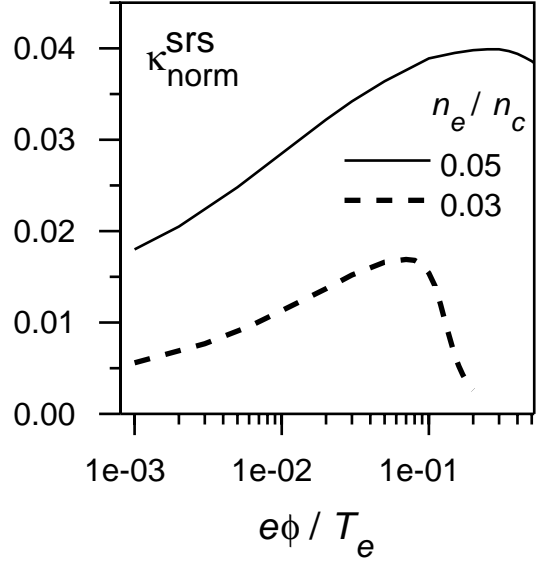

Fig. 3. Maximum (over wavenumber) normalized gain rate as a function of wave amplitude, for $n_e/n_c=0.03$ (dashed), 0.05 (solid), with $T_e=1$keV, and $\nu/\omega_p=0.005$

The actual reflectivity has complex time dependence[22], though apparently statistically stationary, so it is natural to compute the time average reflectivity, $<R>$, and compare the maximum value of $0.5\times\left|d\ln\langle R\rangle/dz\right|$ (the factor of 0.5 because $R$ is proportional to the power of the scattered wave, *i.e.*, amplitude squared) with the upper bound for $\kappa^{srs}$. After 100 ps of evolution from quiescent initial conditions, time averages are accumulated over the next 400ps. Since no low frequency plasma response is allowed in this model, which would otherwise lead to density depletion and subsequent detuning and weakening of the SRS, long time transients are not expected.

The first case considered at 5% critical density, case "A", is for $I$=4E14 W/cm$^2$, so that $(v_{osc}/v_e)^2$=0.036 and therefore $\kappa_{max}^{srs}=4\pi$ x 0.036 x 0.04/$\mu$m =0.018/$\mu$m. The simulation plasma slab extends from $z$=0 to 1200$\mu$m and <$R$> is shown in figure 4 for a portion which contains an interesting feature: log<$R$> has an inflection point at $z\approx170\mu$m, where its rate of change is a maximum, $\left|0.5 \times d\ln\langle R\rangle/dz\right|_{max} = 0.012/\mu m$ (about twice the linear gain rate), compared with the upper bound of 0.018/$\mu$m.

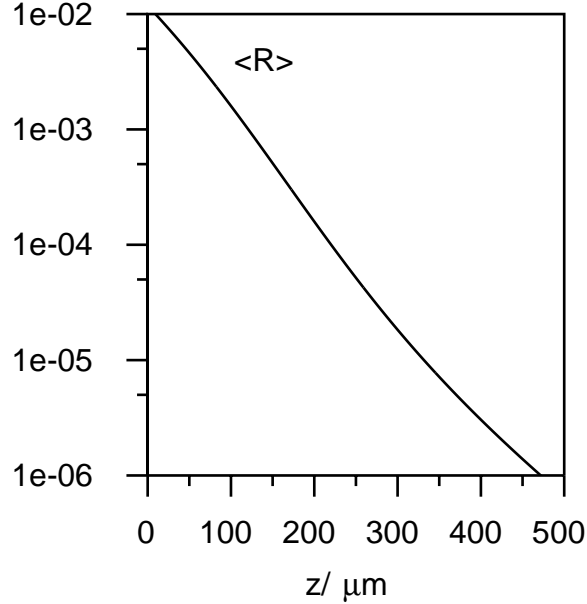

FIG. 4. Mean reflectivity, <$R$>, for $n_e/n_c$=0.05 for 1/2$\mu$m light, $T_e$=1keV, $\nu/\omega_p$=0.005, and $I$=4E14W/cm$^2$.

At this spatial location, $\left\langle\left|e\phi/T_e\right|^2\right\rangle^{1/2} = 0.013$, which is small compared to the value required for $\kappa$ to attain its maximum and <$R$> = 0.0003. However, $R$ and $\phi$ vary exponentially so that even in a small neighborhood, *e.g.*, consider that range of z where the gain rate is within 1% of its maximum, it is found that their values cover a substantial range, $0.0002<R<0.0006$ and $0.01<\phi<0.02$. Therefore, their precise values at the maximum are of no particular significance.

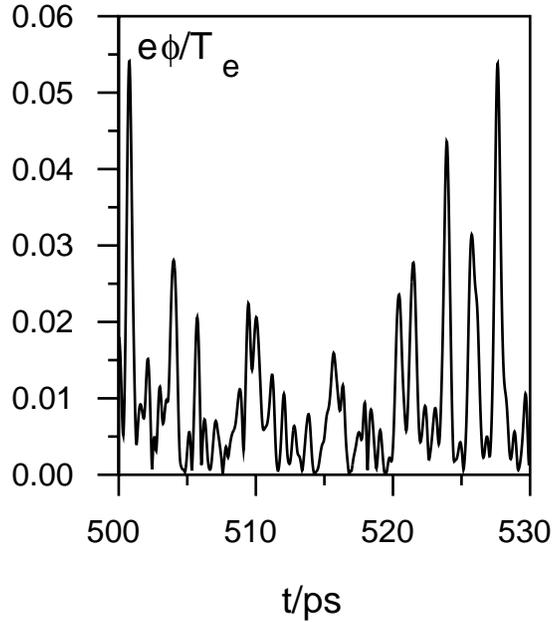

FIG. 5. Amplitude of $\phi$ at $z=170\mu m$, where the average gain rate is a maximum.

$\phi$'s time variation at the inflection point is shown in figure 5. It is remarkable that such a complex process is closely bounded by an estimate that ignores explicit time dependence: the bound exceeds the actual maximum value by only 50%. Part of the reason for this may be interpreted as being due to the insensitivity of $\kappa^{srs}$ to $\phi$, as shown in figure 3, for values of $e\phi/T_e$ greater than about 0.01.

While there is no expectation that $\kappa^{srs}_{max}$ can accurately represent $\kappa^{srs}$ globally, and therefore cannot be used to make a reliable estimate (except in the upper bound sense) for the reflectivity, it is gratifying to note for the above example that although $<R>$ increases by a factor of 100 between the spatial location where $\kappa$ attains its maximum and the left hand boundary of the simulation domain, the gain rate decreases by only 24%.

When $I$ is increased by a factor of 10 to 4E15 W/cm$^2$, length and time scales decrease, and the gain rate bound is still valid, although a bit closer to that based on the time averaged reflectivity: 0.18/$\mu$m *vs* 0.13/$\mu$m and there is little change in the simulation results if the convective term is omitted from (14): instead of 0.13/$\mu$m, the maximum gain rate is reduced to 0.12/$\mu$m. As the remaining cases will be at lower density, where the strong damping approximation is expected to have a greater domain of validity, this convective term will henceforth be dropped.

At lower densities, beyond LOR for the Langmuir wave, there emerges a discrepancy between the upper bound theory and the MCM. The reason for this is discussed in the next section, but first two examples. If the density is reduced to 0.025 of critical, with other parameters the same as the previous example, it is found that $\left(\kappa_{norm}^{srs}\right)_{max}=0.0082$ at $k/k_D=0.50$ with $e\phi/T_e=0.02$ and $\nu/\omega_b=0.07$, so that $\kappa_{max}^{srs}=0.037/\mu$m, while the MCM time average maximum gain rate is 0.025/$\mu$m: so far, so good. But when the density is further reduced to 0.02 of critical, case "B", $\left(\kappa_{norm}^{srs}\right)_{max}=0.0029$ at $k/k_D=0.55$ with $e\phi/T_e=0.003$ and $\nu/\omega_b=0.16$, so that $\kappa_{max}^{srs}=0.013/\mu$m, while the MCM time average maximum gain rate is 0.018/$\mu$m, breaking the putative bound.

## C. Failure of first order mode coupling theory near loss of resonance and an alternative theory—the mean field approximation

The significant omission from equation (14) is the failure to retain higher order time derivatives, since near LOR, the real part of the susceptibility is near a maximum as a function of v, and therefore so is $\varepsilon$ as a function of $\omega$. As given, the time derivative term in equation (14) has the potential for precisely compensating any change in $\text{Re}(\varepsilon/\partial\varepsilon/\partial\omega)$ due to a fluctuation in $\phi$ simply by allowing $\phi$ to evolve at the corresponding frequency, *i.e.*, the first order MCM allows arbitrary amplitude resonant waves. However, as shown in [7], for given $k$ there is a finite value of $\phi$ beyond which waves cannot be resonant[23]. Therefore a model based on the linear frequency interpolation of the dielectric function, such as (14) (equivalently, see (20) below), is unreliable even for small amplitude waves near LOR, *e.g.*, near $k\lambda_D \approx 0.53$.

For yet larger $k\lambda_D$ the use of such a model is more problematic because there is still the possibility of resonance for any $\phi$, while the actual dynamics in this regime is such that resonance is not possible even for small amplitude waves, and the larger $\phi$, the further from resonance is the plasma response. Deep into this regime, trapping decreases the gain rate, contrary to the prediction of the MCM.

This is illustrated in the following two figures that show $\kappa_{norm}^{srs}$, maximized over frequency for each value of $\phi$, as determined by (19) but in which $\varepsilon$ is approximately evaluated as in the MCM:

$$\varepsilon(|\phi|,\omega) = \varepsilon(|\phi|, k_{env}, \omega_{env}) + (\omega - \omega_{env})\partial\varepsilon/\partial\omega. \qquad (20)$$

$\partial\varepsilon/\partial\omega$ is evaluated at $(|\phi|, k_{env}, \omega_{env})$. The strong damping approximation has been invoked since the interest here is in the large $k\lambda_D$ regime. For given plasma parameters, the envelope values of $k$ and $\omega$, $k_{env}$ and $\omega_{env}$, are chosen to maximize $\kappa^{srs}$ for $\phi = 0$. $T_e$ is fixed at 5keV and for the three values of $n_e/n_c$ considered, 0.07, 0.10 and 0.15, the corresponding values of $k_{env}\lambda_D$ are found to be 0.57, 0.45 and 0.33 respectively. $\nu/\omega_p = 0.005$ and the dependence of the results on its value is discussed later. Figure 6 compares the maximum of $\kappa_{norm}^{srs}$ over $\omega$ as computed by the MCM with that computed using the full evaluation of $\varepsilon$ as determined by (12).

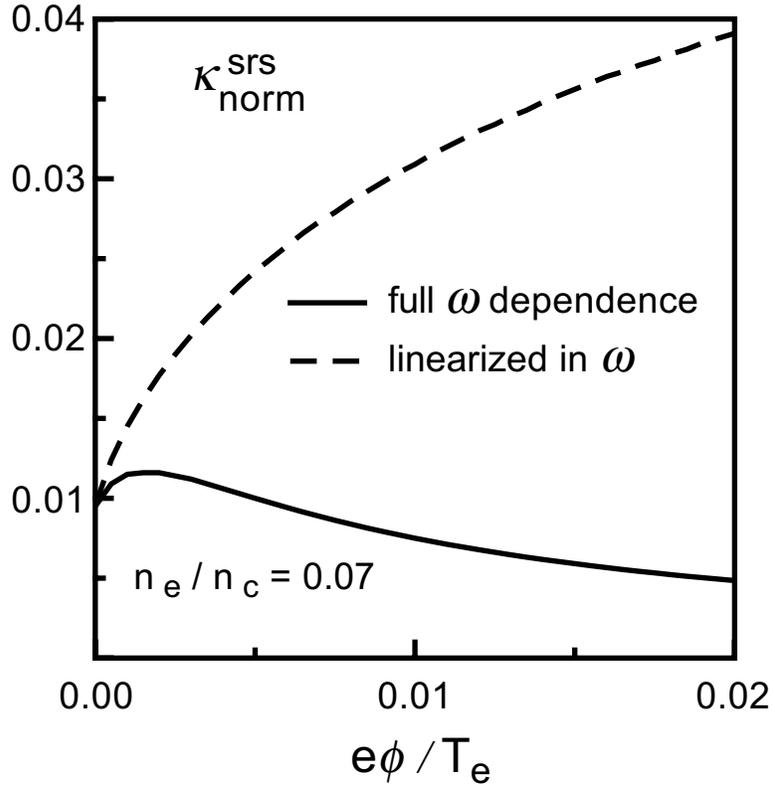

FIG. 6. Maximum normalized gain rate for $n_e/n_c = 0.07$: dashed curve is the mode coupling model result obtained from equations (19) and (20), while the solid curve is obtained from (19) with $\varepsilon$ determined by (12).

The two curves join at $\phi=0$ because of the choice of $k_{env}$. Although a Langmuir resonance is not possible at $k\lambda_D = 0.57$, the solid curve first increases slightly with $\phi$ before decreasing because the plasma is driven near resonance and the cost of being slightly further from resonance due to trapping is offset by the larger decrease in damping due to trapping. The MCM result (dashed curve) departs dramatically from the full theory (solid curve) for small $\phi$: it is too large by a factor of 2 for $\phi=0.0035$.

The next case, $n_e/n_c = 0.1$, is shown in figure 7. Since $k_{env}\lambda_D=0.45$, a Langmuir resonance is possible until $\phi=0.05$, at which point the two theories begin to sharply diverge. For $n_e/n_c = 0.15$, a resonance is possible even for $\phi>0.5$, and over the range of $\phi$ shown in figure 7 the agreement between the two theories is excellent.

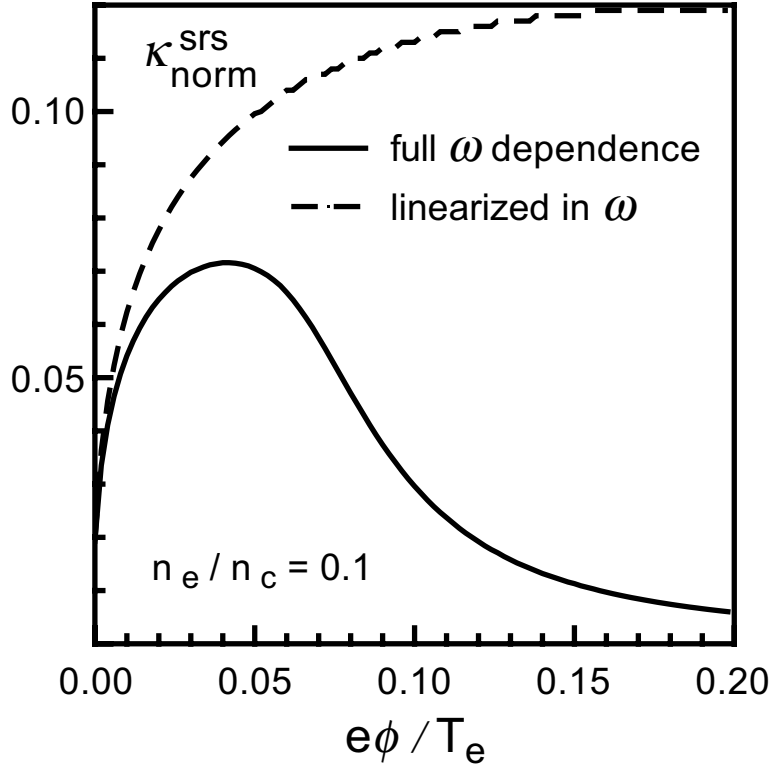

FIG. 7. As in figure 6 except at the higher density, $n_e/n_c = 0.1$

When a resonance is possible, the maximum gain rate varies as $1/\nu$, and is found to be a relatively insensitive function of this parameter otherwise. The solid curve in figure 6 is essentially unchanged when $\nu$ is increased to 0.01, while the MCM result decreases to about 0.02 at $\phi=0.02$: the two theories are in better agreement. For the $n_e/n_c = 0.1$ case, the full theory result is diminished for small $\phi$, when a resonance in possible, but not at $\phi$

=0.2, while the MCM value for $\kappa$ is reduced by a factor of 2 at $\phi=0.2$. Again it is found that the increase of $\nu$ results in better agreement between the theories.

While it is beyond the scope of this paper to study the model obtained by retaining the second order time derivative term, the following model, which is non-perturbative in frequency, is a natural generalization of the first order model, equation (14). Fourier transform (FT) (14) in time, let $\delta\omega$ be the frequency variable, and take the strong damping limit so that the convective derivative is omitted. Multiply through by $\partial\varepsilon/\partial\omega$, in effect deconstructing the formal manipulations that led to (14) in the first place, to obtain

$$\left(\varepsilon + \delta\omega \frac{\partial\varepsilon}{\partial\omega}\right)\hat{\phi}(\delta\omega) \approx \varepsilon(\omega+\delta\omega)\hat{\phi}(\delta\omega) = \hat{\phi}_0(\delta\omega). \tag{21}$$

"^" denotes the FT representation. The transition from (14) to (21) has summed all the $\delta\omega$ corrections, or back in the temporal domain, all time derivatives. The $\phi$ dependence of $\varepsilon$ is evaluated in the mean field approximation (MFA), $\varepsilon(|\phi|) \to \varepsilon(\phi_{rms})$, $\phi_{rms}^2 = \int |\hat{\phi}(\varpi)|^2 d\varpi$, which in practice is approximated by a finite discrete sum. For the MFA to make sense, the actual solution to the dynamic SRS process must be approximately statistically stationary.

For each value of $\omega+\delta\omega$, the wavenumber argument of $\varepsilon$ is chosen according to (17). The actual variation of $k$ is small in practice so that the factor of $k^2$ may be taken as fixed in (16). Equations (13), (16) and (21) imply

$$\left[-i\delta\omega - i\frac{k^2}{8\omega_{srs}}|v_{osc}|^2 + v_{srs}\frac{\partial}{\partial z}\right]\hat{E}_{srs}(\delta\omega) = -\frac{ik^2|v_{osc}|^2}{8\omega_{srs}\varepsilon^*(\omega-\delta\omega,\phi_{rms})}\hat{E}_{srs}(\delta\omega). \tag{22}$$

$|\phi(-\delta\omega)|^2$ can be expressed in terms of $C(\delta\omega) \equiv |E'_{srs}(\delta\omega)|^2$ through the use of (13) and (21). It follows from (22) that

$$v_{srs} \frac{\partial}{\partial z} C(\delta\omega) = \frac{k^2 |v_{osc}|^2 \text{Im}[\varepsilon(\omega-\delta\omega, \phi_{rms})]}{4\omega_{srs} |\varepsilon(\omega-\delta\omega, \phi_{rms})|^2} C(\delta\omega). \tag{23}$$

Since $R \propto \sum_\varpi C(\varpi)$, its logarithmic derivative cannot exceed that of any of its components, and therefore the gain rate upper bound determined by (19) cannot be violated by (22). Equation (22), the MFA model (MFAM), may be viewed as a generalization of the basic MCM which, unlike the latter, is valid in the regime where the daughter SRS Langmuir wave is near or past a loss of resonance. The breadth of the spectrum, $\Delta\omega$, is limited, however, if the dependence of $\varepsilon$ on $\phi_{rms}$ is to be accurately given as if $\phi$ were a coherent wave. The estimate, $\Delta\omega < \omega_b$, is also a constraint on the MCM.

Note that the generalization[24] of (22) to higher dimensions is straightforward through the addition of diffraction, refraction and the dependence of $\varepsilon$ on slow variations in the plasma density, while the generalization of (23) is not because of diffraction. Unless the density fluctuation is slowly varying, it may couple the SRS daughter plasma wave to others. For example, if this wave is unstable to a decay process, then another component to the plasma wave response must be explicitly included.

The validity of the MFAM's bound prediction may be determined by comparison with first principle kinetic simulations. Since $\Xi$ and therefore the gain rate bound explicitly depends on $f_0$, the background distribution function, and since long time, multi-dimensional (multi-D) evolution will lead to its modification[25], this comparison is nontrivial in higher dimensions. Also, quantitative comparisons call for a re-evaluation of $\Xi$ based on transversely localized multi-D Vlasov traveling wave solutions[26].

### D. Comparison of the two models

The MFAM will now be compared with the MCM, starting with case "A". The MCM is expected to be well within its domain of validity since the optimal wavenumber and potential amplitude are found to be $k\lambda_D=0.332$ and $e\phi/T_e=0.24$, while at the nearby wavenumber $k\lambda_D=0.338$, with constraint (17), one finds the maximum value of $\phi$ consistent with a resonance is at $e\phi/T_e=1.04$, both of which greatly exceed the observed fluctuations in $\phi$: at the spatial location of maximum average gain rate, the time series generated by the MCM has $e\phi_{rms}/T_e=0.013$, with a maximum value, over 400ps of observation, of only 0.056. *Both models are found to have the same peak amplitude gain rate* of $0.012/\mu m$, although the MFAM takes longer to get going, attaining its peak value at $z=73\mu m$, which is $100\mu m$ beyond where the MCM attains its maximum[27].

One way to put this spatial shift in perspective is to note that the gain rate only changes by 10% over this $100\mu m$. Another is to study the sensitivity of these results to the thermal noise source. A thermal source term has been added to the rhs of (22) whose magnitude is chosen to give the same level of Thomson scatter as in the MCM. Its frequency content is determined as in the former case: a Lorentzian centered about the linearly SRS matched frequency, and a width determined by linear Landau damping. If the noise source is doubled in the MFAM (*i.e.*, double the Thomson scatter), there is a 50$\mu$m shift in the location of the maximum gain rate towards larger $z$, which brings it closer to the MCM result. On the other hand, if the noise source in the MCM is halved, the location of its peak gain rate is shifted by 60$\mu$m towards smaller $z$, closer to the MFAM result. Therefore the 100$\mu$m shift in the location of peak gain rate, between the MCM and MFAM with the same level of thermal fluctuations, is not viewed as physically significant since it can be compensated for by modest changes in noise levels, a point well worth reconsidering if comparisons are made with particle simulation results.

For case "B", the predictions of the models are qualitatively different. The MFAM yields a maximum amplitude gain rate of 0.01/$\mu$m, which is below the predicted upper bound, 0.013/$\mu$m, and well below the overactive MCM rate of 0.018/$\mu$m. Also, since the MCM is a nonlinear model, it is possible that its maximum gain rate may be even larger when there is a finite BSRS seed.

# IV. EXAMPLES OF BSRS MAXIMUM GAIN RATE FOR NIF AND NOVA RELEVANT PARAMETERS

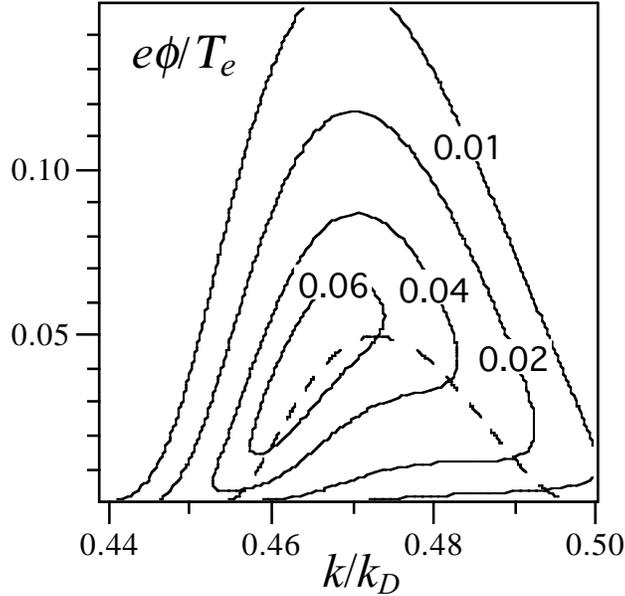

FIG. 8. Contours of normalized gain rate, $\kappa_{norm}^{srs}$, for $\nu/\omega_p$=0.005, $n_e/n_c$=0.1 and $T_e$=5keV.

In figure 8, $\kappa_{norm}^{srs}$ is shown as a function of $k$ and $\phi$ for the case $\nu/\omega_p$=0.005, $n_e/n_c$=0.1 and $T_e$=5keV. The dashed curve is the locus of nonlinear resonance, Re($\varepsilon$)=0, constrained by (17). As the density is varied, for fixed value of the other parameters, the maximum normalized gain rate[28] is shown in figure 9 for $T_e$=2.5 and 5 keV. For example, detailed inspection of the data which generated figure 8 shows that the maximum, $\left(\kappa_{norm}^{srs}\right)_{max}$, is at $k\lambda_D$=0.465, e$\phi$/$T_e$= 0.042 with a value of 0.072, which corresponds to the point on the solid curve in figure 9 at $n_e/n_c$=0.1.

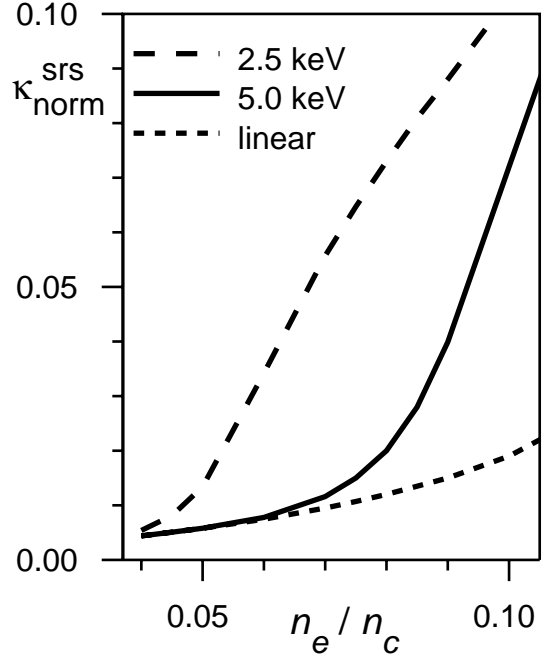

FIG. 9. The maximum normalized BSRS gain rate, for $\nu/\omega_p=0.005$, and $T_e=2.5$ or 5keV. The lower curve, "linear", is obtained by finding the maximum of $\kappa^{srs}_{norm}$ with the additional constraint that $\phi=0$, at 5keV. It joins the nonlinear maximum near $0.07n_c$. At about $0.08n_c$, there is a LOR for the nonlinearly optimal Langmuir wave at 5keV. This may be seen in figure 10 where the optimum[29] value of $k\lambda_D$ is plotted[30].

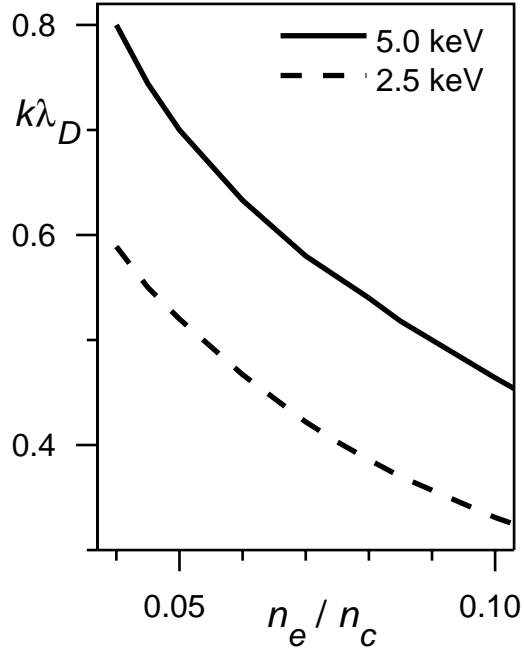

FIG. 10. The optimum plasma wave's wavenumber, for $T_e$ = 2.5 and 5.0 keV. Although there is no possibility of resonance, the departure from resonance is soft compared to the decrease of damping for small $\phi$ at $0.08n_c$, so that the optimum value of $\phi$ is finite and the gain rate is nearly twice the linear value[31]. The optimum value of $\phi$ is shown in figure 11.

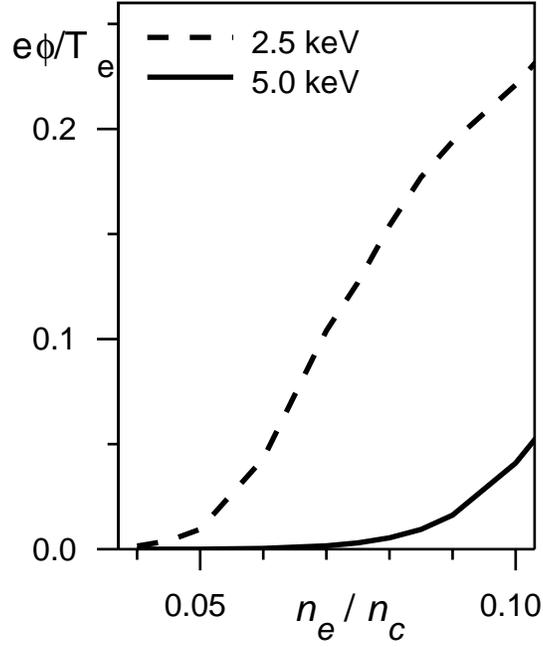

FIG. 11. The optimum plasma wave's amplitude, for $T_e$= 2.5 and 5.0 keV.

At $0.07n_c$ (and, or course, lower densities) however, there is little to gain by allowing for a finite amplitude response at 5keV

While the linear gain rate decreases by a factor of two as the density varies from $0.1n_c$ to $0.07\ n_c$, as seen in figure 9, $\left(\kappa^{srs}_{norm}\right)_{max}$ decreases by a factor of six. Thus, there is a large potential advantage in controlling BSRS by such a change in density, much larger than suggested by linear theory.

Also shown in these figures is the 2.5keV case, corresponding to the regime attained in certain NOVA experiments[32], which attempted to determine a low-density cutoff for BSRS in a hohlraum environment. No cutoff was found, with the average reflectivity a maximum at the lowest density studied, $0.06n_c$. For this lower temperature case, LOR is not until about $0.05n_c$, so that the gain can be enhanced significantly by trapping effects at $0.06n_c$: the ratio of maximum nonlinear gain rate to linear $\approx 4$.

At $0.07n_c$ and 5keV, $\left(\kappa_{norm}^{srs}\right)_{max}$ is 0.0116, which is essentially the linear classic estimate. For $0.35\mu$m laser wavelength and an intensity of $2\times10^{15}$W/cm$^2$, $(v_{osc}/v_e)^2 \approx 0.018$, and therefore $\kappa_{max}^{srs} \approx 0.0037/\mu$m. Since for these plasma parameters, a linear amplitude gain coefficient of about 10 is required to boost thermal fluctuations to order unity, this gain rate would need to be maintained over 2.7mm before BSRS can be significant. This maximum is at $e\phi/T_e=0.0016$, $k\lambda_D=0.581$, $v/v_e=2.3$.

The 2.7mm length scale estimate is upset, however, by the large intensity fluctuations inherent to random phase plate (RPP) optics, where now, *e.g.*, $2\times10^{15}$W/cm$^2$ is the average laser intensity, $<I>$. In particular, a *linear model* of stimulated scatter in the strongly damped regime, for which the field fluctuations are taken to be Gaussian—as it would be in a quiescent plasma—shows that these fluctuations can lead to a reflectivity divergence[33] over a finite plasma slab once $<I>$ exceeds a critical value, $I_c$, which may be qualitatively estimated for a homogeneous plasma by requiring that, for $I=I_c$, a *power* gain coefficient of unity obtains over a speckle length$\approx 7F^2\lambda_0$, or about $150\mu$m for f/8 optics. The value $\kappa_{max}^{srs} \approx 0.0037/\mu$m, in fact, yields just about such a gain exponent.

Quantitative estimates of $I_c$ for BSRS in a hot spot field would be useful since the previous calculation[33] was for BSBS, for which the backscattered light has essentially the same wavelength as the laser, making the latter a more efficient process than BSRS. In addition, propagation over several mm of plasma tends to degrade the laser beam's spatial coherence, *without significantly changing the distribution of hot spot (speckle) intensity[34], if the optic temporal bandwidth is large enough to suppress self-focusing*, effectively shortening the speckle length and thus raising the value of $I_c$. Both these effects raise the critical intensity threshold over the simple estimate of the preceding paragraph.

## IV. LOSS OF RESONANCE FOR ION ACOUSTIC WAVES AND BSBS

When ion dynamics is allowed, it is well known that at low frequencies there are linear ion acoustic modes. Less widely appreciated is the fact that[35] their nonlinear extension to BGK modes has a LOR as various parameters are varied. Besides $k\lambda_D$, there are the ionic composition and electron to ion temperature ratio, $T_e/T_i$, which may be varied. Just as in the case of the Langmuir wave LOR and BSRS, the ion acoustic LOR will separate qualitatively different regimes of BSBS.

### A. Ion acoustic loss of resonance

The nonlinear susceptibility (defined so that $\phi = \phi_0/\varepsilon$ still holds but now the external potential, $\phi_0$, acts on all species) may be expressed as a weighted sum of the previously defined electron susceptibility function, $\Xi$, $\varepsilon = 1 - \Xi_{total}/(k\lambda_D)^2$, with

$$\Xi_{total} = \Xi(v/v_e, e\phi/T_e, \mu) + \sum_i \Theta_i \Xi_i \approx \left[\exp(-e\phi/T_e) - 1\right]/e\phi/T_e + \Xi_{ion}. \tag{24}$$

It has been assumed that $v/v_e \ll 1$, appropriate for the ion acoustic regime, as discussed in II.A, and

$$\Xi_{ion} = \sum_i \Theta_i \Xi_i, \quad \Theta_i = Z_i^2 T_e n_i / T_i n_e, \tag{25}$$

$$\Xi_i = \Xi\left(\frac{v}{v_i}, \frac{Z_i e\phi}{T_i}, \mu_i\right). \tag{26}$$

The values of $i$ run over the number of ion species. $Z_i$ is the i'th species charge state and $n_i$ its average number density. $T_i$, $v_i$ and $\gamma_i$ are the corresponding background ion temperature, thermal speed and trapped ion escape rate respectively, and $\mu_i = \gamma_i / k v_i$. If each $\gamma_i$ is determined by a speckle width thermal ion transit time, so that $\gamma_i \sim v_i$, then all the $\mu_i$ are identical to each other and, at low density, close in value to $\mu$, the dimensionless trapped electron escape rate relevant for BSRS.

In the linear limit, each $\Xi_i$ reduces to $Z'/2$ when $f_0$ is Maxwellian, as in (3), but with argument $v/v_i\sqrt{2}$. Instead of searching for complex frequency roots of $\varepsilon = 0$, the resonances, Re($\varepsilon$)=0, for real $k$ and $\omega$, are sought, as in [35], to find the dispersion relation of small amplitude BGK modes.

It is sufficient to search for LOR in the linear regime to illustrate the rich structure of LOR as a function of the various parameters, so that for this purpose

$$\Xi_{total} = -1 + \sum_i \Theta_i \Xi_0(v/v_i). \tag{27}$$

A fundamental property of $\Xi_0(v)$ (for Maxwellian $f_0$), is that its real part attains a maximum value $\approx 0.285$ at $v \approx 2.13$, and then goes to zero for large v, as in figure 12.

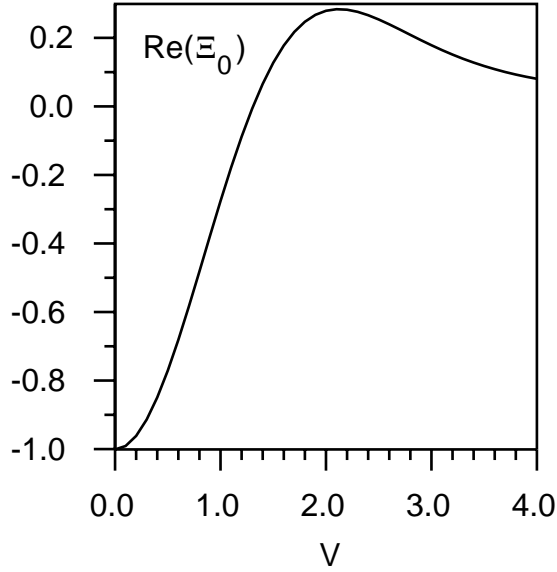

FIG. 12. Real part of $\Xi_0(v) = Z'\left(v/\sqrt{2}\right)/2$ for a Maxwellian background distribution function.

It follows that if a single ion species is to have a chance of a resonance, then[35] $0.285\Theta_i > 1$, or $Z_i T_e/T_i > 1/0.285 \approx 3.5$, since $Z_i n_i = n_e$ is the condition for charge neutrality. If this inequality is satisfied then there are two resonances for small $k\lambda_D$, just as in the pure electron case, but now $\Xi_{total}$ passes through zero at two finite values of v, so that as $k \to 0$, the two resonances have finite phase velocity, *i.e.*, there are two small amplitude BGK acoustic modes. This contrasts with previous[36,37] studies of acoustic solutions to the *linear* dispersion relation that apparently requires two ion species to have the possibility of two weakly damped modes.

## B. Helium-Hydrogen plasma examples

In figure 13 the resonance phase speeds are graphed, normalized to the proton thermal speed, $v_p$, for $T_e/T_i = 5$ and two different Hydrogen fractions, $f_H = n_H/(n_H + n_{He})$, as a function of $k\lambda_D$. The key feature of interest here is the dependence of LOR on $T_e/T_i$ and $f_H$. For example, if $f_H =0.5$ then there is no resonance for $k\lambda_D > 0.8$.

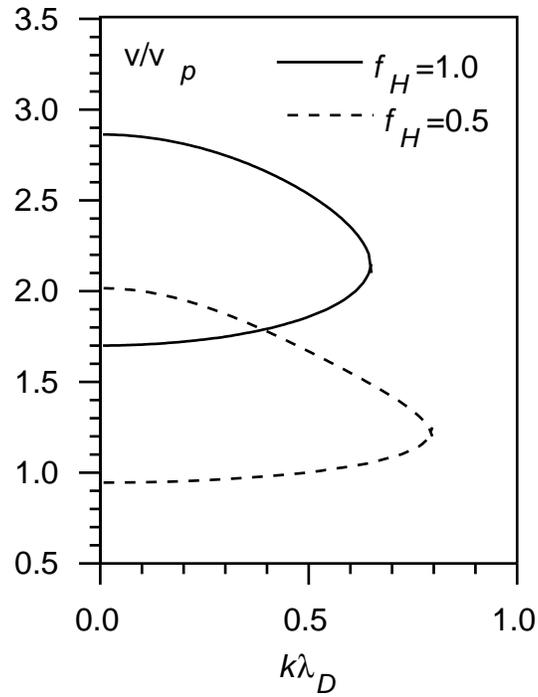

FIG. 13. Normalized ion acoustic phase speeds for $T_e/T_i=5$, parameterized by Hydrogen fraction.

Figure 14 graphs this LOR value of $k\lambda_D$ as a function of $f_H$ and $T_e/T_i$.

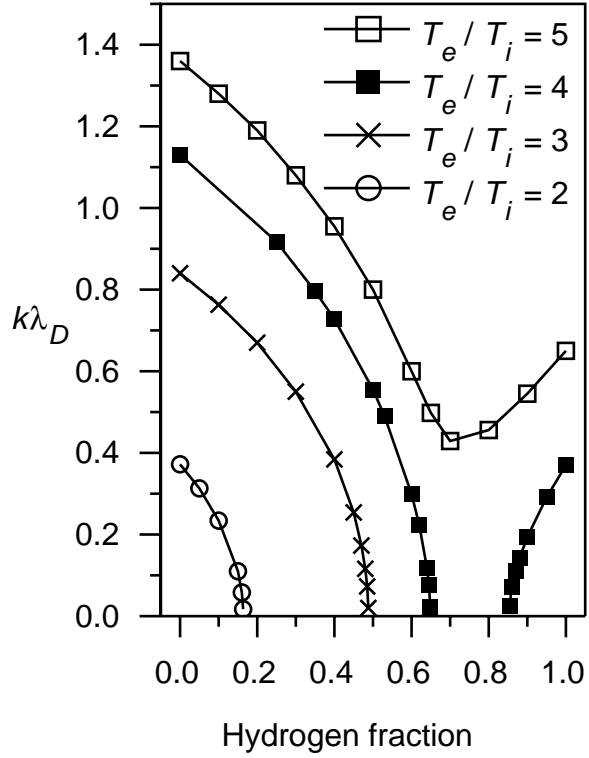

FIG. 14. For given $T_e/T_i$, these curves give the loss of resonance value of $k\lambda_D$. Above such a curve, there is no possibility of an ion acoustic resonance in a Helium-Hydrogen plasma mixture.

For $T_e/T_i=2$, there is no resonance at any wavenumber for $f_H$ greater than about 0.17, while for $T_e/T_i=3$ the cutoff is at about 0.5. If $T_e/T_i=4$, there is a hydrogen fraction gap, from roughly 0.65 to 0.85, in which a resonance is not possible.

Ion acoustic LOR is much harder to come by for a C-H plasma, as seen in figure 15, even for $T_e/T_i=2$. The value of $k\lambda_D$ for LOR is greater than one for $f_H=0.5$, a characteristic value for methane like plasmas.

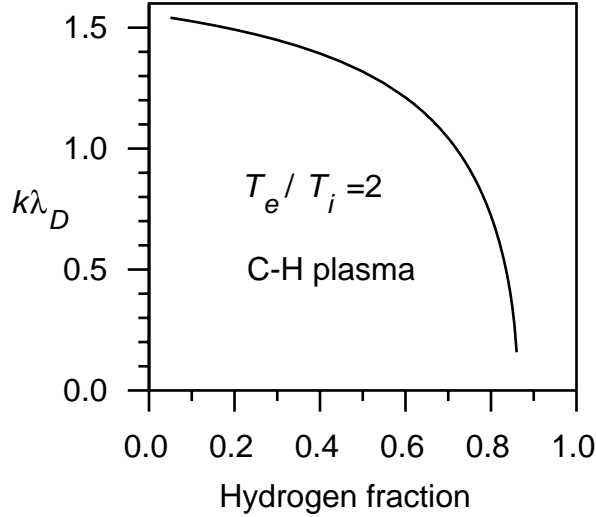

FIG. 15. LOR value of $k\lambda_D$ for a Carbon-Hydrogen mixture.

This has immediate implications for the competition between BSRS and BSBS. Consider a sequence of experiments in which the electron density decreases, *e.g.*, a C-H plasma, as in [32]. As per the discussion in section IV, when BSRS is lost, the value of $k\lambda_D$ for its daughter Langmuir wave is likely to be not much larger[38] than 0.55, and since the wavenumber for the BSBS daughter acoustic wave is at most twice this value, figure 15 implies that the plasma can support a nonlinear ion acoustic mode, even for a temperature ratio as small as 2. It would then be possible for BSBS to exceed linear estimates because of trapping and perhaps win the competition with BSRS. In contrast, a He-H plasma with, *e.g.*, a 60% hydrogen fraction and $T_e/T_i=3$, cannot support a nonlinear acoustic mode, even at zero wavenumber, and BSBS cannot be significantly enhanced by trapping.

## V. SBS IN THE STRONG DAMPING REGIME

Aside from a change in notation, the analysis closely follows that for SRS in section III. The backscattered SBS light, $E_{sbs} = \text{Re}\, E_{sbs} \exp[i(k_{sbs}z - \omega_{sbs}t)]$, has $k_{sbs} \approx -k_0$, and $\omega_{sbs} \approx \omega_0$ since $c_s/c \ll 1$. The ponderomotive potential is given by the analog of (13). However, as it is only the electrons that see this potential, $\phi_0$, the response of the plasma is not given by (18), but may be arrived at by the following variation of a familiar argument. For simplicity, the response of the electron density is linearized, $\delta n_e/n_e = \exp(-e\phi/T_e) - 1 \approx -e(\phi_{int} + \phi_0)/T_e$. Poisson's equation, $(k\lambda_D)^2 e\phi_{int}/T_e = (\delta n_e - \sum_i Z_i \delta n_i)/n_e$, then implies

$$-\frac{e\phi_{int}}{T_e}\left[1 + (k\lambda_D)^2\right] = e\frac{\phi_0}{T_e} + \frac{1}{n_e}\sum_i Z_i \delta n_i. \tag{28}$$

Since the ions only see the internal electric field, for each ion species it follows that,

$$\partial f_i/\partial t + v\frac{\partial f_i}{\partial x} + \frac{eZ_i}{m_i}\frac{\partial \phi_{int}}{\partial x}\frac{\partial f_i}{\partial v} = -v_i(f_i - f_0),$$

and (2) and (25) imply that $\delta n_i/n_i = -e\phi_{int} Z_i \Xi_i/T_i$, with "$\phi$" in (26) replaced by "$\phi_{int}$". The second term on the rhs of (28) may now be evaluated, $\sum_i Z_i \delta n_i/n_e = -\Xi_{ion} e\phi_{int}/T_e$, so that

$$\phi_{int} = \phi_0 \Big/ \left[\Xi_{total} - (k\lambda_D)^2\right]. \tag{29}$$

Poisson's equation then implies

$$\frac{\delta n_e}{n_e} = \frac{e\phi_0}{T_e}\frac{\left[\Xi_{ion} - (k\lambda_D)^2\right]}{(k\lambda_D)^2 - \Xi_{total}}. \tag{30}$$

Equation (13), and the first part of (15) (with "srs"→"sbs"), and (30) then imply

$$\left(\frac{\partial}{\partial t} + v_{sbs}\frac{\partial}{\partial z}\right)E_{sbs} = i\frac{\omega_p^2}{8\omega_{sbs}}\left(\frac{v_{osc}}{v_e}\right)^2\left[\frac{1 - \Xi_{ion}/(k\lambda_D)^2}{\varepsilon}\right]^* E_{sbs}. \tag{31}$$

The convective amplitude spatial gain rate inferred from (31) agrees with previous results[39] when $\phi=0$. Further simplification is gained by using the low density approximation $v_{sbs} \approx -c$,

$$\left(\frac{1}{c}\frac{\partial}{\partial t} - \frac{\partial}{\partial z}\right)E_{sbs} = \frac{ik_0}{8}\frac{n_e}{n_c}\left(\frac{v_{osc}}{v_e}\right)^2 \left[\frac{1-\Xi_{ion}/(k\lambda_D)^2}{\varepsilon}\right]^* E_{sbs}. \tag{32}$$

Therefore the BSBS amplitude convective spatial gain rate is given by

$$\frac{\kappa^{sbs}}{k_0}\left(\frac{v_e}{v_{osc}}\right)^2 = \frac{1}{16 v_{ia}}\frac{n_e}{n_c} \equiv \kappa^{sbs}_{norm}, \tag{33}$$

with

$$\frac{1}{v_{ia}} = -2\frac{\mathrm{Im}(\Xi_{ion})}{\left|(k\lambda_D)^2 + 1 - \Xi_{ion}\right|^2}. \tag{34}$$

$v_{ia}$ may be thought of as a normalized damping rate since in the linear regime with $k\lambda_D \rightarrow 0$, for the case of a single ion species at its acoustic resonance, $v_{ia}$ is the perturbative expression for the ion acoustic amplitude damping rate, normalized to the ion acoustic frequency. As in the case of BSRS, one may seek the maximum of $\kappa^{sbs}$ (minimum of $v_{ia}$) for given laser and plasma parameters, by varying[40] v and $\phi$, with $\Xi_{ion}$ given by (25) and (26).

### A. Damping rate in the linear regime

As a warm up to the nonlinear maximum gain rate determination, consider the maximum linear response. Figure 16 shows its dependence on $T_i/T_e$ and $k\lambda_D$ for $f_H=0.5$, and figure 17 for $f_H=0.7$, in a Helium-Hydrogen plasma. Although the linear response is similar in these two cases, they are both shown by way of contrast with the nonlinear response, discussed in V.B.

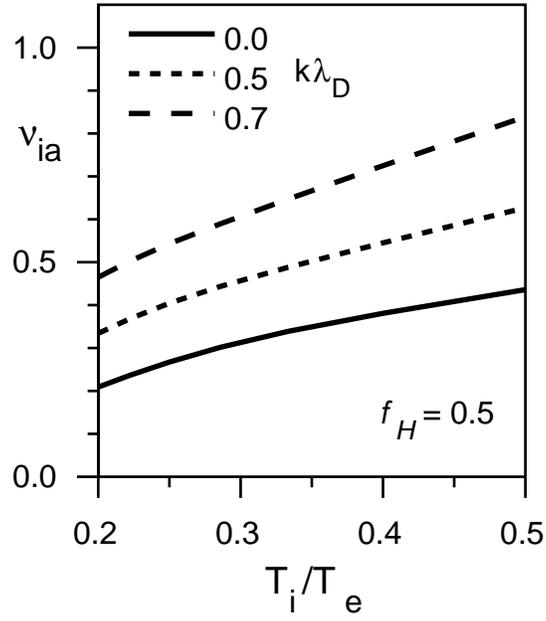

Fig. 16. Normalized damping rate, which determines the gain rate via equation (33), for He-H plasma with equal numbers of H and He atoms, with $k\lambda_D$ as parameter, as determined by (34), for $\phi=0$.

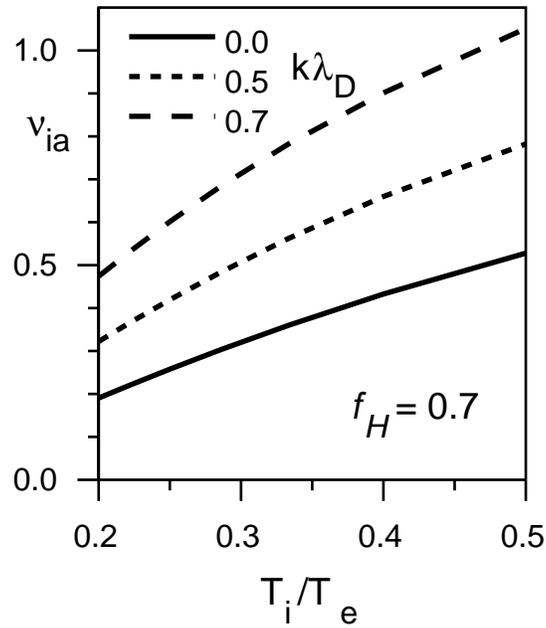

Fig. 17. Normalized damping rate for He-H plasma with H number fraction=0.7.

## B. Maximum response in the nonlinear regime for a Helium-Hydrogen plasma

Since a NIF hohlraum plasma environment is rich in possibilities, with time dependent electron temperature, electron density, laser intensity, $T_e/T_i$ and variable hydrogen fraction, the following examples are chosen to illustrate different regimes of behavior, not "representative" behavior. For all examples, $\mu=0.01$. Sensitivity of the results to this choice is discussed later. Other experiments involving plasma created in a He-H gas jet have been proposed[41] in which $T_e/T_i$ is estimated to be as large as 8 for $T_e \approx 1$kev.

The first example has $T_e=1$keV, with $T_e/T_i=4$, for which there is a qualitative difference between the $f_H=0.5$ and $f_H=0.7$ cases, as seen in figure 18 which shows the normalized maximum BSBS amplitude gain rate[42], $\kappa_{norm}^{sbs}$, given by (33).

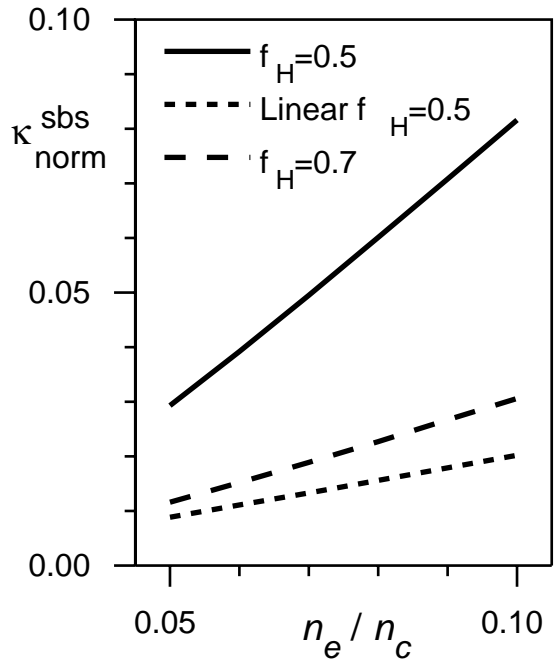

Fig. 18. Normalized maximum BSBS amplitude gain rate for $T_e=1$keV, $T_e/T_i=4$ and, $\mu=0.01$. The bottom curve, "Linear", is the maximum for $\phi=0$ and $f_H=0.5$.

There is little difference between the maximum linear response for $f_H = 0.5$ (shown) and 0.7 (not shown). Figure 19 shows the corresponding optimum values of $\phi$.

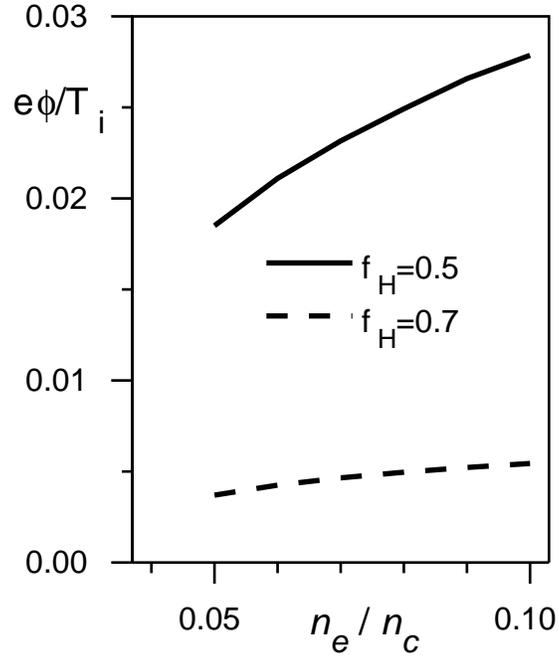

FIG. 19. Optimum value of $\phi$ for $T_e=1\text{keV}, T_e/T_i=4$ and , $\mu=0.01$.

Resonance is not possible for $f_H=0.7$, whereas it is possible over this range of densities for $f_H=0.5$ (see figure 14). This is consistent with the fact that the ratio of the responses is much greater than would be inferred from the ratio of acoustic damping coefficients computed in the linear regime, as in figures 16 and 17. For this temperature and range of densities, $k\lambda_D$ varies between roughly 0.3 and 0.4.

Figure 20 shows the maximum normalized gain rate for different values of $T_e/T_i$ and $n_e/n_c=0.07$ for the 1keV case. It may be seen that $f_H=0.7$ is about the best choice[43] for the purpose of minimizing the maximum gain rate, though it remains a sensitive function of temperature ratio.

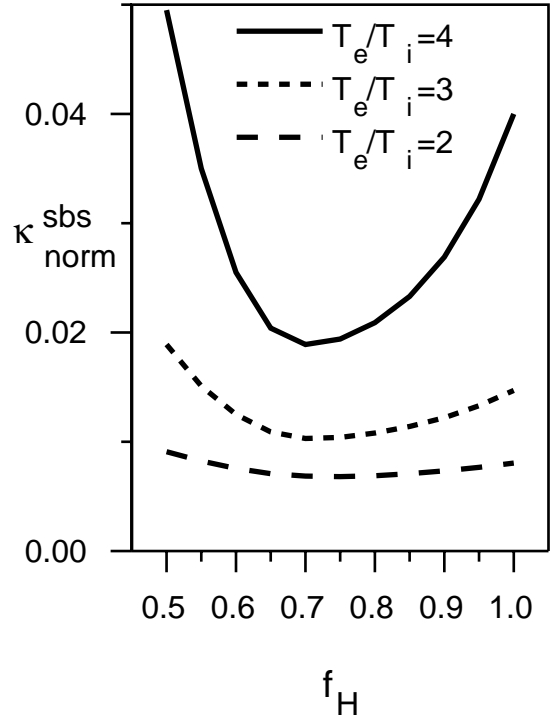

FIG. 20. Normalized maximum BSBS amplitude gain rate for $T_e$=1keV, $n_e/n_c$=0.07, and $\mu$=0.01

For reference the elementary graph of $2k_0\lambda_D$, where $k_0$ is the laser wavenumber in the plasma, and $2k_0 \approx k$, the ion acoustic wavenumber for BSBS, is shown in figure 21 for $T_e$=5keV.

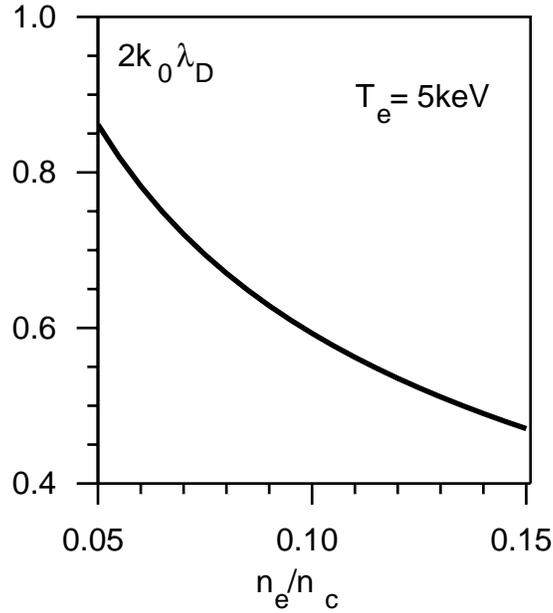

Fig. 21. $k\lambda_D = 2k_0\lambda_D$ for the BSBS daughter ion acoustic wave.

The next example is at the higher temperature, $T_e$=5keV, with $T_e/T_i$=2. Maximum gain rate results are shown in figure 22. In contrast to the first example, the gain rate is an insensitive function of hydrogen fraction. The linearly and nonlinearly maximum gain rates are the same for the $f_H$ =0.7 case and almost the same for the $f_H$ =0.5 case which benefits slightly (a fraction of a percent) when finite $\phi$ is allowed at $n_e/n_c$=0.09 and 0.10. A resonance is not possible for any of these higher $T_e$ cases. Note the factor of 10 change in scale between figure 18 and figure 22. Now consider a more particular numerical example which may be deemed appropriate to later times of a NIF laser pulse: the parameters of figure 22, with $I$=2x10$^{15}$W/cm$^2$, 0.07 critical density, and $f_H$=0.7, yields $\kappa_{max}^{sbs} \approx (2\pi/0.35\mu m)$ x 0.018 x 0.004=0.0013/$\mu$m, which requires 7.7mm to come up from thermal fluctuations. That distance is halved by taking into account intensity fluctuations, because these provide at least a factor of 2 enhancement over the gain based on <$I$> since the backscattered light may be phase conjugate with the incident laser, even if the laser's spatial coherence is degraded due to propagation over a large plasma length. This estimate does not vary much with $\mu$: it changes less than 1% as $\mu$ varies from 0.005 to 0.02, as expected since this case is far from resonance.

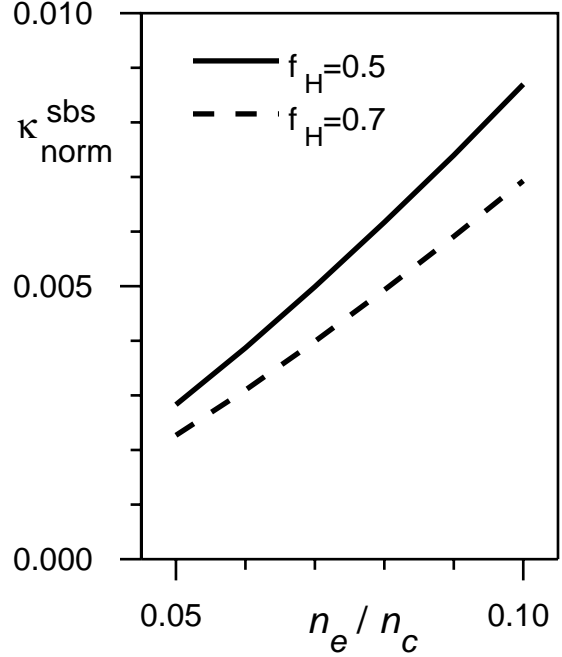

FIG. 22. Normalized maximum BSBS amplitude gain rate for $T_e$=5keV, $T_e/T_i$ =2 and, $\mu$=0.01.

## VI. SUMMARY AND FUTURE DIRECTIONS

The simplest predictor as to the importance of trapping in stimulated scatter is the possibility of a plasma wave resonance. If a resonance is possible for small amplitude waves, which may easily be discerned from the real part of the linear dielectric function, then as the wave amplitude, $\phi$, is increased there is a decrease of damping, persistence of resonance and the stimulated scatter spatial gain rate, $\kappa$, increases until $\phi$ gets large enough so that there is a loss of resonance (LOR), and then $\kappa$ tends to decrease (but see figure 6 for an example which shows that trapping may increase $\kappa$ close to LOR[44]). The maximum value that $\kappa$ attains before decreasing due to LOR is the main subject of this paper. It is calculated with a previously derived[7] perturbative expression for the nonlinear dielectric function.

Electron trapping effects are your ally for controlling BSRS in the high temperature regime, $T_e$=5keV, at $n_e$=0.07$n_c$ (except for a small increase in $\kappa$ for small $\phi$ as in figure 6) and conversely not your ally at 0.1$n_c$. Linear theory indicates only a factor of 2 difference in BSRS gain rate at these two densities, while trapping can cause as much as a factor of 6. Below 0.07$n_c$, the BSRS gain rate decreases slowly with density (see figure 9). Proton trapping suppresses BSBS at both densities if $T_e/T_i<3$ and the hydrogen fraction is at least 0.5.

Since the first order mode coupling model (MCM), in particular equation (14), always allows for a resonant plasma response, such a model is unreliable when applied to the calculation of BSBS in a He-H plasma in the NIF high temperature regime in which an ion acoustic resonance cannot occur. Depending on the choice of hohlraum plasma density, and the corresponding presence or absence of electron plasma wave resonances, such a model is also inadequate for BSRS. While in principle, the first order theory may be generalized to include a 2$^{nd}$ order time derivative, thereby regaining information about LOR, an alternative mean field approximation model was introduced whose validity does not require a plasma resonance. Even if $\varepsilon$ is determined by the perturbative expansion of $\Xi$, as in (5), (6) and (12), whose validity is not limited to small wavenumbers, and whose implied frequency shift is non-perturbative, once there is LOR it, as well as any other first order in time MCM, will fail.

The magnitude of the gain estimates for the NIF examples is not surprising. Reference [39] reports maximum amplitude gains of about 10 for both BSRS and BSBS, although quantitative comparison is not possible since in that work the linear gain is calculated by integrating over spatially varying intensity, temperature and density profiles relevant to the basic NIF hohlraum design. What is new here is the conclusion that at the peak of the laser pulse, the estimate for BSBS cannot be increased by trapping at $0.1n_c$ (or lower), while that for BSRS may be significantly increased by trapping at $0.1n_c$, but not at $0.07n_c$.

Within the framework of stimulated scattering theory presented here, there are several apparent avenues of research that should yield physically more relevant estimates. First the spatial coherence of the laser light is degraded after propagating through several mm of plasma. This phenomenon, its effect on the critical intensity and on the mean gain rate for average intensity below critical for the case of BSRS, need to be quantitatively modeled. Second, if plasma parameters are such that daughter plasma waves are resonant, then a better estimate for the damping due to the escape of trapped electrons (or protons for the case of BSBS) in a 3D model should be obtained to remove the uncertainty of the dimensional analysis estimate used in the 1D model. Third, if the hohlraum plasma fill is at an electron density, say $0.1n_c$, at which both electron plasma waves are resonant at 5keV, then one may need to calculate the nonlinear dielectric function with a better approximation than would be obtained by merely superposing their effects.

## ACKNOWLEDGEMENTS

This work was performed under the auspices of the U. S. Department of Energy by the Los Alamos National Laboratory under contract No. W-7405-Eng-36. I thank Bruce Cohen and Ildar Gabitov for helpful discussions, and Paul A Bradley for NIF simulation results.

Figure captions

FIG. 1. The normalized escape rate of trapped electrons, $\mu=\nu/kv_e$, must be well below the ordinate of this graph for trapping to effect a significant reduction in damping.

Fig. 2. Various approximations to the imaginary part of the nonlinear susceptibility, $\Xi$, for $v/v_e=3.0$ and $\mu=0.01$.

Fig. 3. Maximum (over wavenumber) normalized gain rate as a function of wave amplitude, for $n_e/n_c=0.03$ (dashed), 0.05 (solid), with $T_e=1$keV, and $\nu/\omega_p=0.005$

FIG. 4. Mean reflectivity, $<R>$, for $n_e/n_c=0.05$ for $1/2\mu$m light, $T_e=1$keV, $\nu/\omega_p=0.005$, and $I=4E14$W/cm$^2$.

FIG. 5. Amplitude of $\phi$ at $z=170\mu$m, where the average gain rate is a maximum.

FIG. 6. Maximum normalized gain rate for $n_e/n_c=0.07$: dashed curve is the mode coupling model result obtained from equations (19) and (20), while the solid curve is obtained from (19) and $\varepsilon$ is determined by (12).

FIG. 7. As in figure 6 except at the higher density, $n_e/n_c=0.1$

FIG. 8. Contours of normalized gain rate, $\kappa_{norm}^{srs}$, for $\nu/\omega_p=0.005$, $n_e/n_c=0.1$ and $T_e=5$keV.

FIG. 9. The maximum normalized BSRS gain rate, for $\nu/\omega_p=0.005$, and $T_e=2.5$ or 5keV.

FIG. 10. The optimum plasma wave's wavenumber, for $T_e=2.5$ and 5.0 keV.

FIG. 11. The optimum plasma wave's amplitude, for $T_e=2.5$ and 5.0 keV.

FIG. 12. Real part of $\Xi_0(v)=Z'(v/\sqrt{2})/2$ for a Maxwellian background distribution function.

FIG. 13. Normalized ion acoustic phase speeds for $T_e/T_i=5$, parameterized by Hydrogen fraction.

FIG. 14. For given $T_e/T_i$, these curves give the loss of resonance value of $k\lambda_D$. Above such a curve, there is no possibility of an ion acoustic resonance in a Helium-Hydrogen plasma mixture.

FIG. 15. LOR value of $k\lambda_D$ for a Carbon-Hydrogen mixture.

Fig. 16. Normalized damping rate for He-H plasma with equal numbers of H and He atoms, with $k\lambda_D$ as parameter, as determined by (34), for $\phi=0$.

Fig. 17. Normalized damping rate, which determines the gain rate via equation (33), for He-H plasma with H number fraction=0.7.

Fig. 18. Normalized maximum BSBS amplitude gain rate for $T_e$=1keV, $T_e/T_i$=4 and, $\mu$=0.01. The bottom curve, "Linear", is the maximum for $\phi=0$ and $f_H=0.5$.

FIG. 19. Optimum value of $\phi$ for $T_e$=1keV, $T_e/T_i$=4 and , $\mu$=0.01.

FIG. 20. Normalized maximum BSBS amplitude gain rate for $T_e$=5keV, $T_e/T_i$=2 and, $\mu$=0.01.

Fig. 21. $k\lambda_D=2k_0\lambda_D$ for the BSBS daughter ion acoustic wave.

FIG. 22. Normalized maximum BSBS amplitude gain rate for $T_e$=1keV, $n_e/n_c$=0.07, and $\mu$=0.01

[15] B. I. Cohen and A. N. Kaufman, Phys. Fluids **21**, 404 (1978).

[16] For example, it may be obtained from equation (7.10) in W. L. Kruer, *The Physics of Laser Plasma Interactions*, 1st edition, edited by D. Pines (Addison-Wesley, New York, 1988), Chap. 7, page 76, by using the envelope representations for the light and plasma waves. The equation for the laser envelope field is omitted for simplicity since it is not needed to obtain the gain rate upper bound. Collisional absorption is ignored.

[17] A qualitatively similar model has been studied by H. X. Vu, D. F. DuBois, and B. Bezzerides, Phys. of Plasmas **9**, 1745 (2002).

[18] For numerical results, the exact, finite density, light wave dispersion relation is used.

[19] For Maxwellian $f_0$, the Langmuir branch of BGK modes is more responsive than the electron acoustic branch, so that for the purpose of optimizing the gain rate, it is BSRS which wins.

[20] If the imaginary part is retained, then short wavelength fluctuations are unstable, which violates the smoothly varying envelope ansatz.

[21] When diffraction of the laser light is included, then in regions of space where there is constructive interference, speckles or intensity hot spots, the laser intensity may attain values large compared to its average, <*I*> and upper bound estimates based on a model with uniform intensity at that average value may be wildly off the mark. For example, the absolute instability threshold may be locally exceeded in a collection of hot spots, but not exceeded in the corresponding uniform case. More importantly, the distinction between models with and without diffraction is most dramatic when <*I*> exceed its critical value. See the discussion in section IV.

[22] This has been seen in particle simulations, H. X. Vu, D. F. DuBois, and B. Bezzerides, Phys. Rev. Lett. **86**, 4306 (2001).

[23] The related phenomenon of wave breaking gives a similar estimate, for example, see T. P. Coffey, Phys. Fluids **14**, 1402 (1971) and W. L. Kruer, in *The Physics of Laser Plasma Interactions*, **1**st edition, edited by D. Pines (Addison-Wesley, New York, 1988), Chap. 9, p.104.

[24] The MFAM may be extended to include pump depletion by evaluating the mode coupling term in the equation for $E_0$, which is proportional to $E_{srs}\phi$, as $\sum E'_{srs}(\delta\omega)\phi'(-\delta\omega)$.

[25] In a multi-speckle environment, the inter-speckle distribution function, which serves as the "background" for any given speckle, will differ from that at infinity.

[26] It has recently been shown that there are no such traveling wave solutions, Li-Jen Chen, David Thouless and Jian-Ming Tang, Bull. of the APS **47**, No. 9, page 291 (2002), but if the previous estimate for the loss of trapped electrons is valid, then large enough amplitude waves will appear as traveling wave solutions for many bounce periods.

[27] Since the simulation region is 1200$\mu$m long, the MFAM takes about 1100$\mu$m to come up from thermal fluctuations, while the MCM only takes about 1000$\mu$m.

[28] For perfect consistency of notation, the ordinate of figure 9 should be encased by parenthesis with the subscript "max", but in the interest of a lighter notation this is not done. The figure caption is unambiguous.

[29] In the sense that it yields the maximum gain rate.

[30] Recall that if it exceeds 0.53, resonance is not possible.

[31] Vu, DuBois and Bezzerides, reference [22], have also noted a significant enhancement of gain rate above the linear estimate for k$\lambda_D$ as large as 0.55.

[32] Juan C. Fernández, J. A. Cobble, D. S. Montgomery, M. D. Wilke, and B. B. Afeyan, Phys. Plasmas **7**, 3743 (2000).

[33] Harvey A. Rose and D. F. DuBois, Phys. Rev Lett., **72** 2883 (1994).

[34] Andrew J. Schmitt and Bedros B. Afeyan, Phys. Plasmas **5**, 503 (1998). Also R. L. Berger and E. A. Williams, private communication (1999).

[35] M. Buchanan and J. Dorning, Phys. Rev. E **52**, 3015 (1995).

[36] H. X. Vu, J. M. Wallace, and B. Bezzerides, Phys. Plasmas **1**, 3542 (1994).

[37] E. A. Williams, R. L. Berger, R. P. Drake, A. M. Rubenchik, B. S. Bauer, D. D. Meyerhofer, A. C. Gaeris, and T. W. Johnston, Phys. Plasmas **2**, 129 (1995).

[38] This is a qualitative statement. The laser intensity could be made large enough so that BSRS acting in isolation could be significant even for $k\lambda_D \gg 0.55$.

[39] For example, see the integrand for the spatially integrated *power* gain, G, which is a factor of 2 larger than the amplitude gain rate reported here, on page 2030 of B. J. MacGowan, B. B. Afeyan, C. A. Back, R. L. Berger, G. Bonnaud, M. Casanova, B. I. Cohen, D. E. Desenne, D. F. DuBois, A. G. Dulieu, K. G. Estabrook, J. C. Fernandez, S. H. Glenzer, D. E. Hinkel, T. B. Kaiser, D. H. Kalantar, R. L. Kauffman, R. K. Kirkwood, W. L. Kruer, A. B. Langdon, B. F. Lasinski, D. S. Montgomery, J. D. Moody, D. H. Munro, L. V. Powers, H. A. Rose, C. Rousseaux, R. E. Turner, B. H. Wilde, S. C. Wilks, Phys. Plasmas **3**, 2029 (1996). Also see E. A. Williams, *Phys. Plasmas* **3**, 2029 (1996), with $\chi_e$ replaced by $1/(k\lambda_D)^2$.

[40] As is well known, because of the disparity between the ion acoustic and scattered light speed at low density, the ion acoustic wave has a wavenumber k≈2$k_0$, and it is more convenient to characterize its response by its frequency, or phase velocity v, rather than *k*, unlike the SRS case.

[41] Private communication, David S. Montgomery, 2002.

[42] For perfect consistency of notation, the ordinate of figure 18 should be encased by parenthesis with the subscript "max", but in the interest of a lighter notation this is not done. The figure caption is unambiguous.

[43] Based on linear dispersion relation analysis, reference [36], it has been found $f_H$=0.5 is optimal for minimizing the SBS response, although the nonlinear analysis presented here shows that there may be a significant advantage in choosing $f_H$=0.7.

[44] If the calculation shown in figure 6 is repeated for $n_e/n_c = 0.08$, it is found that the gain rate increases until e$\phi$/$T_e$ = 0.006 at which point it is nearly twice the value at $\phi$=0 even though $k_{env}\lambda_D$ for this case is 0.52, on the verge of LOR.